\documentstyle[12pt]{article}

\newcommand{\newsection}{
\setcounter{equation}{0}
\section}
\newcommand{\tr}[1]{\,{\rm tr}\,#1\,}

\def\eop{\vspace*{\fill}\pagebreak}
\def\be{\begin{equation}}
\def\ee{\end{equation}}
\def\bea{\begin{eqnarray}}
\def\eea{\end{eqnarray}}

\def\l{\lambda}

\newcommand{\dd}[1]{{\partial \over \partial #1}}

\title{{\bf \mbox{} \\The Spatial  Dynamics in Kazakov--Migdal Model.}
\vspace{.5cm}}
\author{{\bf K. Zarembo}
\date{ }
\vspace{.5cm} \\
{\it Steklov Mathematical Institute} \\
{\it Vavilov st.42, GSP-1, 117966 Moscow, Russia}}

\begin{document}

\maketitle

\vspace{-9.6cm}

\begin{flushright}
SMI--06--92 \\
hep-th/9301032 \\ December, 1992
\end{flushright}

\vspace{10.8cm}

\begin{abstract}
The spatially inhomogeneous  large $N$ solutions to  Kazakov--Migdal
model are analyzed. The set of nonlinear differential equations  is
derived in the continuum limit. In one dimensional case these
equations has a  natural interpretation in terms of  the dynamics
of a Fermi gas.  The multidimensional case seems to be inconsistent
because of  its instability related to the  collapse of eigenvalues
of the scalar field.

\end{abstract}

\eop

\newsection{Introduction.}

A while ago, Kazakov and  Migdal \cite{KM} proposed a lattice
gauge model induced by a heavy scalar field in  the adjoint  representation
of $SU(N)$. The action of  this  model is the usual gauge  invariant action
without Yang--Mills term:
\be
S=-\sum_{x}N\,\tr{\left[U_{0}\left(\Phi\left(x\right)\right)-\frac{1}{2}
\sum_{\nu =-D}^{D}\Phi (x)\Omega_{\nu}(x)\Phi (x+\nu )\Omega_{\nu}
^{\dagger}(x)\right]}.
\label{1.1}
\ee
Although  there is some problems with the  induction of a physical
QCD \cite{KSW,KhM,Mak,MigMix} it is interesting to investigate
the continuum limit of this model. In this paper we study a
semiclassical dynamics of the  density of  eigenvalues of matrix
$\Phi (x)$ :
\be
\rho (\lambda,x) ={1 \over N} tr{\delta\left(\lambda-
\Phi(x)\right)}.
\label{1.2}
\ee
An integral  equations for $\rho (\lambda,x)$ can be obtained
using the technique developed  by Migdal  \cite{MigEx,MigMix}.
When the lattice spacing goes to zero a continuum  limit
can be constructed about the critical  potential  $U_{cr}\left(\Phi
\right)=D\Phi^2$. It is  worth mentioning that the physical mass
of scalar particles  goes  to zero in  lattice units as  the local
limit  is approaching in contrast to what is necessary  for reproducing
of  QCD.

A set of nonlinear  differential equations is  obtained in the
continuum limit. In $D=1$ case  these equations has a
hydrodynamical interpretation and the translationally
 invariant solution is stable. When $D>1$ the situation is qualitatively
the other. The spectrum  of  excitations about the spatially
homogeneous solution is  always tachyonic. So the
continuum limit  in $D>1$ case seems to be physically  unacceptable
because of  its  instability.

\section{Saddle Point Equations.}

\setcounter{equation}{0}
Due to gauge invariance all the matrices $\Phi(x)$ can  be
diagonalized by gauge transformation. So,  fixing the  diagonal
gauge and integrating over link variables one obtains  the
effective  action depending upon $\Phi_{i}(x)$  -- the eigenvalues
of  $\Phi(x)$. In the  large $N$  limit WKB approximation becomes
exact. The semiclassical equation of motion reads  as follows
\be
-U_{0}'(\lambda)+2W(\lambda,x)+\sum_{\nu=-D}^{D}F_{\nu}(\lambda,x)=0,
\label{2.1}
\ee
$\lambda$ varying along the support of $\rho(\lambda,x)$. The second
term comes from gauge fixing determinant:
\be
W(\l,x)=\wp\int d\xi\, \frac{\rho(\xi,x)}{\lambda-\xi}.
\label{2.2}
\ee

$F_{\nu}(\lambda,x)$ is the logarithmic derivative of the
Itzykson--Zuber integral:
\be
F_{\nu}(\lambda,x)=\lim_{N \rightarrow  \infty}\left.\frac{1}{N}
\dd{\Phi_{i}(x)}
\ln\int D\Omega \,\hbox{e}^{\,N\tr{\Phi(x)\Omega\Phi(x+\nu)
\Omega^{\dagger}}}\right|_{\Phi_{i}(x)=\lambda}.
\label{2.3}
\ee
Of course, it depends only upon the eigenvalue   densities of
$\Phi(x)$ and $\Phi(x+\nu)$. Using the Schwinger--Dyson
equations  for Itzykson--Zuber integral Migdal obtained
the following dispersion relation determining $F_{\nu}(\lambda,x)$
in terms of $\rho(\lambda,x)$ and $\rho(\lambda,x+\nu)$ \cite{MigEx}:
\be
W(\lambda,x+\nu)=\int \frac{d\xi}{\pi}\,\arctan  \frac{\pi
\rho(\xi,x)}{\lambda-F_{\nu}(\xi,x)-W(\xi,x)}.
\label{2.4}
\ee

In  spatially  homogeneous  case the equations  (\ref{2.1}),
(\ref{2.2}) and (\ref{2.4}) were analyzed  by Migdal \cite{MigEx} in
some detail and were solved exactly for the  Gaussian potential
by Gross \cite{Gr}. In the present paper we are interested
in a  spatial dynamics of the eigenvalue  density.

\section{The Quadratic Potential.}

\setcounter{equation}{0}
Before  studying the general equations it is instructive to
consider a more simple model with  purely quadratic potential
$U_{0}(\Phi)=1/2m_{0}^{2}\Phi^{2}$. The saddle point equations
can be simplified  in this case due to the observation \cite{Gr}
that translationally invariant semi-circular distribution of
eigenvalues solves (\ref{2.1}), (\ref{2.2}), (\ref{2.4}).
The  semi-circular ansatz is useful  in the case with  a spatial
fluctuations too:
\be
\rho(\l,x)=\sqrt{\mu\left(  x\right)-\frac{1}{4}\mu^{2}\left( x
\right) \l^{2}},~~~~F_{\nu}=\frac{1}{2}f_{\nu}(x)\l.
\label{3.1}
\ee
Substituting  (\ref{3.1})  into (\ref{2.2}) and (\ref{2.4}) and
doing the  integrals we express $W(\l,x)$ and $f_{\nu}(x)$ in terms
of  $\mu(x)$. All the  functions $U_{0}'(\l)$, $W(\l,x)$ and
$F_{\nu}(\l,x)$ are proportional to $\l$ with coeffitient of
proportionality depending on  the  $\mu(x)$ only, so the $\l$-dependence
in the equation (\ref{2.1}) can be eliminated, that gives
\be
\mu(x)=m_{0}^{2}-\frac{1}{2}\sum_{\nu=-D}^{D}\left[\sqrt{\mu^{2}(x)+
4\frac{\mu(x)}{\mu(x+\nu)}}-\mu(x)\right].
\label{3.2}
\ee
This equation has  translationally invariant solution \cite{Gr}
\be
\mu_{\pm}=\frac{m_{0}^{2}(D-1)\pm D\sqrt{m_{0}^{4}-4(2D-1)}}
{2D-1}.
\label{3.3}
\ee
It is interesting that (\ref{3.2}) has no strongly fluctuating
antiferromagnetic solutions.

Now we are going to take the continuum limit of the equation  (\ref{3.2}),
 so we  rescale  $m_{0}^{2}=m^{2}a^{2}+2D$. At that moment a difference
between $D=1$ and $D>1$ cases appears. Really, in the former case
$\mu_{+}$ vanishes  as the lattice spacing goes to zero. From
(\ref{3.3}) we see that $\mu_{+}$ scales as  $2ma$, so rescaling
$\mu(x)\rightarrow \mu(x)a$  and expanding (\ref{3.2})  in $a$ up to
the second order we get  for $\sqrt{\langle\frac{1}{N}
\tr\Phi^{2}(x)\rangle}=\mu^{-1/2}
(x)\equiv  \phi(t)$, $t=ix$  the equation
\be
\ddot{\phi}-m^{2}\phi+\frac{1}{4\phi^{3}}=0.
\label{3.4}
\ee
This  describes  an  oscillations with doubled   frequency $2m$ about
the static  solution $\phi=(2m)^{-1/2}$.

In  multidimensional case $\mu_{-}$ vanishes in  the local limit
and scales as $-m^{2}a^{2}/(D-1)$, so after rescaling $\mu(x)\rightarrow
\mu(x)a^{2}$ we  obtain  from (\ref{3.2}) the following equation
(in the Minkowski space):
\be
\Box\phi-m^{2}\phi-\frac{D-1}{\phi}=0.
\label{3.5}
\ee
The  effective potential for $\phi$ is unbounded from below, so
the collapse  of eigenvalues taking place in the naive  (unregularized)
continuum limit is inevitable. Translationally invariant solution
$\phi=\sqrt{-(D-1)/m^{2}}$ corresponds to a maximum of the effective
potential, so it is unstable.

\section{The Continuum Limit  for an Arbitrary Potential.}

\setcounter{equation}{0}
Let us consider the one dimensional case first.  The canonical
scaling dimensions  in (\ref{2.1}), (\ref{2.2}) and (\ref{2.4}) are
recovered by the substitution $\l\rightarrow\l a^{-1/2}$,
$\xi \rightarrow \xi a^{-1/2}$, $U'_{0}(\l) \rightarrow 2\l a^{1/2}+
aU'(\l)$, $\rho(\xi,x) \rightarrow  \rho(\xi,x)a^{1/2}$. For $F_{\pm}(\l,x)$
we can write  $F_{\pm}(\l,x)=\l a^{-1/2}+\left[v_{\pm}(\l,x)-W(\l,x)
\right] a^{1/2}+G_{\pm}(\l,x)a^{3/2}$. The  first term  in this expression
is chosen so  that (\ref{2.4}) becomes the  identity at the vanishing
lattice spacing. The first and the second order terms in (\ref{2.4})
reads
\be
\int d\xi \,\frac{\rho(\xi,x)v_{\pm}(\xi,x)}{(\l-\xi)^{2}}=
\pm \dd{x}  W(\l,x)
\label{4.1}
\ee
\be
\int  d\xi \,\frac{\rho(\xi,x)}{(\l-\xi)^{2}} \left[G_{\pm}(\xi,x)+
\frac{v_{\pm}^{2}(\xi,x)-{1 \over 3} \pi^{2}\rho^{2}(\xi,x)}{\l-\xi}\right]=
\frac{1}{2}{\partial^{2} \over \partial x^{2}} W(\l,x).
\label{4.2}
\ee
 From (\ref{2.1}) we have
\be
v_{+}(\l,x)+v_{-}(\l,x)=0
\label{4.3}
\ee
\be
U'(\l)=G_{+}(\l,x)+G_{-}(\l,x).
\label{4.4}
\ee
The functions $v_{+}(\xi,x)$ and  $v_{-}(\xi,x)$ differs only by sign,
so we denote $v(\xi,t)=-iv_{+}(\xi,$ $x)=iv_{-}(\xi,x)$, where
 $t=ix$.
After integration by parts  in the  l.h.s.  of (\ref{4.1}) we obtain
the following equation for $v$
\be
\frac{\partial\rho}{\partial t}+\dd{\xi} (\rho v)=0.
\label{4.5}
\ee

It is well known, that $\rho(\xi,t)$ may be interpreted as the density of a
Fermi gas in an external potential \cite{BIPZ}. So the equation (\ref{4.5})
has a natural interpretation as the hydrodynamical continuity
condition, $v$ being the velocity of a Fermi gas flow. The equation of
motion for it is obtained as  follows. First, differentiate  (\ref{4.5})
in  $t$, substitute the result in (\ref{4.2}) and  integrate by parts
in the r.h.s. Consequently  integrating by  parts the second term
in the square brakets. The last step is the  elimination of  $G_{\pm}$
by  adding  of the two equations (\ref{4.2}) and using (\ref{4.4}).
The result reads
\be
\dd{t} (\rho v)+\dd{\xi}\left[\rho \left(U+v^{2}+\frac{1}{3}\pi^{2}
\rho^{2}\right)\right]-U\frac{\partial \rho}{\partial \xi}=0.
\label{4.6}
\ee

The present consideration is a generalization of a  method used by
Gross  \cite{Gr} to reproduce the well known fermionic solution \cite{BIPZ}
to one dimensional matrix model:
\be
\rho_{0}(\xi)=\frac{1}{\pi}\sqrt{2E-2U(\xi)}.
\label{4.7}
\ee
This is just the static solution to (\ref{4.5}), (\ref{4.6}). Constant $E$,
the Fermi level, is determined by the normilization condition  for
$\rho_{0}$.

In the acoustical approximation -- $\rho(\xi,t)=\rho_{0}(\xi)+u(\xi,t)$,
$\left| u(\xi,t)\right| \ll \rho_{0}(\xi)$,
\be
\frac{\partial^{2}u}{\partial t^{2}}-\dd{\xi}\left[\pi \rho_{0} \dd{\xi}
\left(\pi \rho_{0} u \right)\right]=0.
\label{4.8}
\ee
The local velocity  of sound,  $\pi  \rho_{0}(\xi)$, is equal
to  the Fermi  momentum,  as  one might expect.

Now let us turn to the multidimensional case.  From the analysis of  the
Gaussian potential we  learned that the density of eigenvalues scales as
$a^{1}$. So   we write  $\xi  \rightarrow \xi a^{-1}$,
$\l \rightarrow  \l  a^{-1}$, $\rho(\xi,x) \rightarrow  \rho(\xi,x) a$,
$U'_{0}(\l) \rightarrow 2D\l a^{-1}+U'(\l)$, $F_{\nu}(\l,x)=\l a^{-1}+
v_{\nu}(\l,x)+\left[G_{\nu}(\l,x)-W(\l,x)\right] a$. It can be verified
directly that this is the only  way to obtain the sensible continuum
limit of (\ref{2.1}),  (\ref{2.2}) and (\ref{2.4}). All the steps in a
derivation  of the equations for $\rho$ and $v_{\nu}$ are the same as
in the one dimensional case. After the Wick rotation $x^{D}\rightarrow
-ix^{0}$, $v_{D}\rightarrow iv_{0}$ we get
\be
\frac{\partial \rho}{\partial x^{\nu}}+\dd{\xi}\left(\rho v_{\nu}\right)=0
\label{4.9}
\ee
\be
\dd{x^{\nu}}\left(\rho v^{\nu}\right)+\rho \left[U'+2(D-1)W\right]+
\dd{\xi}\left(\rho v^{2}\right)=0.
\label{4.10}
\ee
For the Gaussian potential  the substitution of (\ref{3.1}) with
$v_{\nu}(\xi,x)$  linear in $\xi$  reduces these equations to  (\ref{3.5}).

Translationally invariant solution to (\ref{4.9}), (\ref{4.10}) is
determined by the following condition
\be
\wp \int d\l \,\frac{\rho_{0}(\l)}{\xi-\l}=-\frac{U'(\xi)}{2(D-1)}.
\label{4.11}
\ee
It coincides with the saddle  point equation for one matrix model with
a potential $-U(\xi)/(D-1)$.  The generic solution to it is \cite{BIPZ,Dav}
\be
\rho_{0}(\xi)=\frac{1}{2\pi}\sqrt{Q(\xi)-\frac{1}{(D-1)^{2}}\left[U'(\xi)
\right]^{2}},
\label{4.12}
\ee
where $Q(\xi)$ is a polinomial of degree less than that of  $U'(\xi)$.
 However, this solution is unstable.  The reason is that the  r.h.s.  of
(\ref{4.11}) has a negative sign, so the eigenvalues of $\Phi(x)$ are
accumulated not in a  minimum, but in a maximum  of the potential. The
instability of the translationally invariant solution, of course, can  be
demonstrated explicitly by linearization of (\ref{4.9}), (\ref{4.10})
near $\rho_{0}(\xi)$. The linearized equation reads
\be
\Box u +2(D-1)\dd{\xi}\left[\rho_{0}\,\wp \int  d\l \,\frac{u(\l,x)}{\xi-\l}
\right]=0.
\label{4.13}
\ee
Substitution $u(\xi,x)=\hbox{e}^{-ik_{\nu}x^{\nu}}u_{0}(\xi)$ leads to
the following eigenvalue problem
\be
Af(\xi)\equiv -2(D-1)\,\wp \int d\l \, \frac{\left[\rho_{0}(\l)f(\l)\right]'}
{\xi-\l}=k^{2}f(\xi).
\label{4.14}
\ee
There $f(\l)=1/\rho_{0}(\l)\,\int_{0}^{\l}  d\xi \,u_{0}(\xi)$.
Operator $A$  is self-adjoint with respect to the scalar product
$(f_{1},f_{2})=\int d\l \, \rho_{0}(\l)f_{1}^{*}(\l) f_{2}(\l)$,
all its eigenvalues being negative, because
\be
(f_{1},Af_{2})=-(D-1)\int d\xi d\l  \,\frac{\left[\rho_{0}(\xi)f_{1}^{*}(\xi)-
\rho_{0}(\l)f_{1}^{*}(\l)\right]
\left[\rho_{0}(\xi)f_{2}(\xi)-\rho_{0}(\l)f_{2}(\l)\right]}{(\xi-\l)^{2}}.
\label{4.15}
\ee
So all the spectrum of  excitations  is tachyonic.

\section{Conclusions.}

\setcounter{equation}{0}
In multidimensional case we does   not find such fine physical picture as
in one dimension. Translationally invariant solution to the semiclassical
equations of  motion corresponds to a maximum of the potential, the
fluctuations about it are unstable. Although we  can not rule out the
existance of a more complicated stable  vacuum, we do not see the physical
reasons for this possibility. So it seems that without the valuable
changes, like an inclusion of fermions  \cite{MigMix,KhMF}, Kazakov-Migdal
model is unstable in  the  continuum limit.

\section{Acknowledgements.}

I would like to  thank  L.Chekhov  for introducing me to  this domain
and  for useful discussions.

\end{document}